\documentclass[%
reprint,
superscriptaddress,
amsmath,
amssymb,
aps,
prb,
]{revtex4-2}

\usepackage{mathtools}
\usepackage{graphicx}%
\usepackage{dcolumn}%
\usepackage{bm}%
\usepackage{lipsum}
\usepackage[utf8]{inputenc}
\usepackage[T1]{fontenc}
\usepackage{etoolbox}
\usepackage{orcidlink}
\usepackage{physics}

\preprint{AIP/123-QED}

\usepackage{graphicx}
\usepackage{dcolumn}
\usepackage{bm}
\usepackage{subfigure}
\usepackage{multirow}
\usepackage{soul}
\usepackage[normalem]{ulem}
\usepackage{xcolor}
\usepackage{kantlipsum}
\usepackage[version=4]{mhchem}
\usepackage{textcomp}

\usepackage{booktabs}
\newcolumntype{C}{>{$}c<{$}}
\AtBeginDocument{
\heavyrulewidth=.08em
\lightrulewidth=.05em
\cmidrulewidth=.03em
\belowrulesep=.65ex
\belowbottomsep=0pt
\aboverulesep=.4ex
\abovetopsep=0pt
\cmidrulesep=\doublerulesep
\cmidrulekern=.5em
\defaultaddspace=.5em
}
\usepackage{siunitx}
\usepackage{multirow}

\definecolor{amber}{rgb}{1,0.49,0}

\newcommand{\editor}[2]{%
  \expandafter\newcommand\csname #1note\endcsname[1]{%
    \textcolor{#2}{(\textbf{#1:} ##1)}}%
  \expandafter\newcommand\csname #1\endcsname[1]{%
    \textcolor{#2}{##1}}%
  \expandafter\newcommand\csname #1cancel\endcsname[1]{%
    \textcolor{#2}{\sout{##1}}}%
  \expandafter\newcommand\csname #1change\endcsname[2]{%
    \textcolor{#2}{\sout{##1} ##2}}%
  \newenvironment{#1text}{\color{#2}}{\color{black}}
}

\newcommand{\ee}{\mathrm{e}}

\definecolor{verde}{rgb}{0.,0.6,0}
\editor{PP}{verde}
\editor{resub}{red}
\editor{FG}{tangerine}

\usepackage{mathrsfs}
\usepackage{placeins}

\begin{abstract}
    Accessing the thermal transport properties of glasses is a major issue for the design of production strategies of glass industry, as well as for the plethora of applications and devices where glasses are employed. From the computational standpoint, the chemical and morphological complexity of glasses calls for atomistic simulations where the interatomic potentials are able to capture the variety of local environments, composition, and (dis)order that typically characterize glassy phases. Machine-learning potentials (MLPs) are emerging as a valid alternative to computationally expensive \textit{ab initio} simulations, inevitably run on very small samples which cannot account for disorder at different scales, as well as to empirical force fields, fast but often reliable only in a narrow portion of the thermodynamic and composition phase diagrams. In this article, we make the point on the use of MLPs to compute the thermal conductivity of glasses, through a review of recent theoretical and computational tools and a series of numerical applications on vitreous silica and vitreous silicon, both pure and intercalated with lithium.
\end{abstract}

\begin{document}

\title{Thermal transport of glasses via machine learning driven simulations}

\author{Paolo Pegolo\,\orcidlink{0000-0003-1491-8229}}
\affiliation{%
 SISSA---Scuola Internazionale Superiore di Studi Avanzati, Trieste
}
\author{Federico Grasselli\,\orcidlink{0000-0003-4284-0094}}
\affiliation{%
COSMO---Laboratory of Computational Science and Modeling, IMX, \'Ecole Polytechnique F\'ed\'erale de Lausanne, 1015 Lausanne, Switzerland
}
\date{\today}
\maketitle

The pursuit of improving the thermal conductivity properties of amorphous solids is central to contemporary materials science and engineering~\citep{mauro2014grand}. Glasses, characterized by their lack of crystalline order, possess unique attributes that make them invaluable across a wide range of applications.
One of the prominent features of this class of materials is a inherently low thermal conductivity, a result of their disordered structure. This property is useful in various fields, such as aerospace engineering~\citep{kotz2017three,hu2020predicting}, electronics~\citep{pasquarello1998interface}, and pharmaceutical industries~\citep{niu2018molecular}.
In contrast, specific industrial applications demand a nuanced balance in the thermal properties of glasses. 
For example, nuclear reactors and nuclear weapon decommissioning generate radioactive waste~\citep{ewing2015long-term} that must be safely stored for exceptionally long time. This waste can be solidified through vitrification, preventing accidental radionuclide release thanks to the amorphous structure of glasses, which provides radiation protection and outstanding chemical durability, thus enabling thousands of years of safe storage~\citep{ojovan2019introduction}. Here, effective heat management is crucial, as high thermal conductivity enhances vitrification efficiency, influencing melt rate and glass homogeneity~\citep{sugawara2014high,kim2015effect}. Moreover, in long-term storage, elevated heat conductivity rapidly dissipates decay-generated heat, avoiding issues like high-temperature-assisted crystallization, porosity, and cracks~\citep{matzke1993thermal}. 
During the last decades, a huge effort has been put forward to access the structural and thermodynamical properties of the glasses employed for nuclear waste vitrification, mainly (boro)silicates, from both the experimental~\citep{kim2015effect,matzke1993thermal,sugawara2014high,kim2017temperature} and the computational sides~\citep{pedesseau2015first,pedone2022interatomic,pallini2023comparison,bertani2023effects,sorensen2021thermal,sorensen2019boron,sorensen2022thermal}. Nevertheless, a microscopic description of thermal conduction in these materials beyond the celebrated Cahill-Pohl model~\citep{cahill1988lattice}, as a function of temperature and composition, is still missing.
The lack of computational studies on thermal conduction is shared by another important application, namely solid-state batteries, where several designs leverage amorphous solid electrolytes~\citep{manthiram2017lithium,zhao2020designing,janek2023challenges,fujita2023structural,landry2023unveiling}. 
Here, the glassy phase is also characterized by a diffusive species (like Li$^+$ or Na$^+$ ions) which poses further challenges for the microscopic simulation of heat transport, since lattice methods cannot be formally applied due to the lack of well defined positions of mechanical equilibrium.

Machine-learning (ML) is an increasingly popular tool in materials modeling due to its ability to train on extensive datasets precise models able to match a wide array of target properties. One route is to leverage datasets of mechanical and functional glass properties to discover new materials~\citep{mauro2016accelerating,onbacsli2020mechanical}, or to predict end-properties such as solubilities~\citep{brauer2007solubility}, dissolution rates~\citep{anoopkrishnan2018predicting}, or transition temperatures~\citep{cassar2018predicting}, to name a few~\citep{liu2021machine}.
By taking a microscopic approach, one can build surrogate models able of \textit{ab initio} interatomic potentials to drive simulations to sample all kinds of properties of amorphous materials that can be accessed by atomistic simulations~\citep{sosso2012thermal,Deringer2017,sosso2018understanding,allen1989thermal,guo2022artificial, Sivaraman_2020,deringer2021origins,paruzzo2018chemicalshifts,islam2021amorphous,brorsson2022efficient,liu2023unraveling,xie2023uncertainty,langer2023heat}.

In this article, we provide an overview of the microscopic theory of heat transport in glasses. Our focus is on methods that effectively utilize ML potentials (MLPs) for computing thermal conductivity. We explore two primary approaches: lattice dynamics, suitable for solids at temperatures significantly below their melting point, enabling the incorporation of quantum-mechanical effects in heat transport; and equilibrium molecular dynamics (EMD) simulations, a versatile tool for sampling material properties, which however is limited in its ability to account for the aforementioned quantum effects. 

\section{Materials and Methods}

\subsection{Thermal transport in glasses}

Heat transport is characterized by the value of the thermal conductivity, $\kappa$, whose linear-response value is given by the Green-Kubo (GK) formula~\citep{green1952markoff,kubo1957statstical1,kubo1957statistical2,baroni2020heat}

\begin{align}\label{eq:gk kappa}
    \kappa = \frac{1}{3 \Omega k_\mathrm{B} T^2} \int_0^\infty \expval{\mathbf{J}_q (t) \cdot \mathbf{J}_q (0)} \dd{t},    
\end{align}
where $\Omega$ is the system's volume, $k_\mathrm{B}$ the Boltzmann constant, $T$ the temperature, and $\mathbf{J}_q$ the heat flux. The factor $1/3$ comes from the assumption of isotropy. 
The temperature-dependent behavior of $\kappa$ in glasses exhibits three distinctive, universally recognized patterns~\citep{beltukov2013ioffe}. At extremely low temperatures, specifically when $T \lesssim 2\,$K, the predominant scattering mechanism involves quantum tunneling between various local minima within the glass energy landscape, resulting in $\kappa \sim T^2$~\citep{phillips1987two, buchenau1992interaction, lubchenko2003origin}. As the temperature reaches a few tens of kelvins, thermal conductivity increases and eventually reaches a plateau value. Despite the absence of a firmly established theoretical consensus in the literature, this phenomenon appears to be linked to the transition from a regime dominated by quantum processes to one where propagating waves are scattered by random disorder~\citep{buchenau1992interaction, lubchenko2003origin, schirmacher2006thermal, beltukov2013ioffe}. Beyond the plateau, the behavior of the thermal conductivity is governed by the anharmonic decay of normal modes~\citep{isaeva2019modeling,simoncelli2019unified}, and $\kappa$ starts increasing again until it saturates to its high-temperature value. We focus here on the latter range of temperatures, which is relevant for applications and can be investigated by means of atomistic simulations.

There are different techniques for computing $\kappa$. Equilibrium methods require the calculation of the heat flux appearing in Eq.~\eqref{eq:gk kappa}~\citep{baroni2020heat}, or the energy density~\citep{drigo2023heat}. Equivalently, one can compute the energy flux along with other relevant mass fluxes, as discussed below. A versatile approach for obtaining these fluxes is through EMD simulations, which allow to sample the energy flux at all orders within anharmonic perturbation theory. A downside of EMD simulations is that it is currently impossible to sample the energy flux including nuclear quantum effects~\citep{habershon2013ring}, despite extensive efforts are being made in this direction [see, e.g. \cite{sutherland2021nuclear,wang2023quantum,siciliano2023wigner} and citations therein]. In solids, an alternative approach lies in exploiting lattice dynamics, at the cost of neglecting high-order anharmonic interactions. At temperatures well below the melting point, atomic nuclei undergo small oscillations around well-defined equilibrium positions. This behavior allows us to represent the dynamics with harmonic normal modes, which, in crystals, are described as phonon quasiparticles. In the harmonic approximation, the normal modes feature infinite lifetimes, resulting in infinite thermal conductivity, regardless of the presence of (harmonic) perturbations, such as disorder~\citep{fiorentino2023unearthing}. However, when anharmonic interactions are appropriately accounted for, e.g., through perturbation theory~\citep{isaeva2019modeling, simoncelli2019unified}, they induce temperature-dependent frequency shifts and broadenings, impacting phonon lifetimes and ultimately determining the thermal conductivity of materials. 

Using lattice dynamics to express the heat flux in Eq.~\eqref{eq:gk kappa} in terms of phonon creation and annihilation operators, one can compute the GK formula for $\kappa$ in the quasi-harmonic Green-Kubo (QHGK) approximation~\citep{isaeva2019modeling,fiorentino2023green} as

\begin{align}\label{eq:kappa qhgk}
    \kappa = \frac{1}{3\Omega} \sum_{\mu\nu} C_{\mu\nu} \abs{v_{\mu\nu}}^2 \frac{\gamma_\mu+\gamma_\nu}{(\omega_\mu - \omega_\nu)^2 + (\gamma_\mu + \gamma_\nu)^2},
\end{align}
where $\omega_\mu$ and $\gamma_\mu$ are the angular frequency and anharmonic linewidth of the $\mu$th normal mode, respectively; $v_{\mu\nu}$ is a generalized velocity matrix, and

\begin{align}\label{eq:two-mode heat capacity}
    C_{\mu\nu}=\frac{\hbar^2\omega_\nu\omega_\mu}{T}\frac{n(\omega_\nu)-n(\omega_\mu)}{\hbar(\omega_\mu-\omega_\nu)}
\end{align}
is a generalized two-mode isochoric heat capacity, $n(\omega)=[\ee^{\hbar \omega /(k_\mathrm{B} T)}-1]^{-1}$ being the Bose-Einstein (BE) occupation function.

The QHGK approach employs two interconnected approximations~\citep{fiorentino2023green, caldarelli2022manybody}. The first is the \emph{dressed bubble} approximation, where four-point correlation functions among phonon creation and annihilation operators are factorized into products of two-point correlation functions, neglecting vertex corrections~\citep{fiorentino2023green}. This implies that normal modes decay independently, each interacting with a common heat bath. The second approximation, termed \emph{Markovian}, disregards memory effects in the heat bath-normal mode interaction~\citep{fiorentino2023green}. The combination of these approximations leads to four-point correlation functions being expressed using single-body \emph{greater} Green's functions, denoted by ${g^>_{\mu}(t)=-\mathrm{i}(n_{\mu}+1)\mathrm{e}^{-\mathrm{i}\omega_{\mu}t -\gamma_{\mu}|t|}}$. The quasi-harmonic hypothesis requires ${\gamma_{\mu}^2/\omega_{\mu}^2\ll 1}$, implying that only nearly-degenerate modes with ${|\omega_{\mu}-\omega_{\nu}|\lesssim \gamma_{\mu}+\gamma_{\nu}}$ significantly contribute to heat conductivity.
The QHGK approximation works on crystals and glasses alike, reducing to the result of the Boltzmann Transport Equation in the former case, and---at the expense of neglecting anharmonic effects---to the Allen-Feldman (AF) model of harmonic disordered solids in the latter~\citep{isaeva2019modeling, barbalinardo2020efficient, fiorentino2023green}. Neglecting anharmonic effects is no trivial matter---it remarkably transforms the finite bulk thermal conductivity of a glass into an infinite quantity, thereby emphasizing the AF calculations' qualitative accuracy driven solely by size effects~\citep{fiorentino2023unearthing,fiorentino2023hydrodynamic}.

If compared with crystals, glasses present an additional complexity in numerical computations due to their inherent aperiodic nature~\citep{fiorentino2023hydrodynamic}. The customary application of periodic boundary conditions (PBC) to finite simulation cells, from which quantities in Eq.~\eqref{eq:kappa qhgk} are derived, requires cautious consideration of size effects. While the physical symmetry of crystals allows for calculations on a fine mesh in reciprocal space, in glasses the same approach is not feasible without introducing spurious contributions to the thermal conductivity~\citep{fiorentino2023hydrodynamic}. Nonetheless, size extrapolation remains possible by leveraging the asymptotic Debye expression of the thermal conductivity of propagons---the low-frequency normal modes in glasses characterized by wave-like properties, including well-defined dispersion with small broadening~\citep{allen1999diffusons,fiorentino2023hydrodynamic}.

Using lattice dynamics requires to be able to compute second and third-order interatomic force constants, i.e., the second and third partial derivatives of the potential energy with respect to nuclear coordinates at equilibrium~\citep{barbalinardo2020efficient}.  This can be achieved through finite differences~\citep{narasimhan1991anharmonic,tang2010anharmonicity} or perturbation theory~\citep{paulatto2013anharmonic}, with the former method being more common. For large systems (i.e., containing tens of thousands of atoms) calculating interatomic force constants is computationally feasible using empirical force fields, but becomes impractical when pursued \emph{ab initio} due to the $N_\mathrm{atoms}^2$ scaling of second and $N_\mathrm{atoms}^3$ scaling of third-order force constants. In principle, MLPs offer a partial remedy, significantly expediting interatomic force computations.
Regrettably, practical implementation is often spoiled by pronounced numerical noise in interatomic force constants, impeding the computation of anharmonic linewidths through Fermi's golden rule~\citep{barbalinardo2020efficient}, as it will be showcased below by a toy example involving vitreous silica. Inaccurate anharmonic linewidths severly hinder the computation of heat conductivities, thereby rendering the entire lattice dynamical workflow unfeasible. 
This concern is particularly relevant at low frequencies~\citep{bruns2022comment}, even for MLPs with good performances within the medium to high-frequency range. A potential strategy to mitigate this issue entails substituting finite-difference third-order derivatives with analytical ones, exploiting the differentiabilty of MLP descriptors through, e.g., automatic differentiation~\citep{langer2023heat}.

Therefore, while the challenge of deriving interatomic force constants from MLPs is under scrutiny, their accuracy in reproducing \emph{ab initio} results can be leveraged in MD simulations through the GK equation, Eq.~\eqref{eq:gk kappa}.

\subsection{Thermal conductivity from molecular dynamics simulations}\label{sec:kappa gk from md}

The GK formula requires an expression for the heat flux. Equivalently, one can combine the \emph{energy} flux with all the independent mass fluxes of the chemical species in the material~\citep{baroni2020heat,bertossa2019theory}. The latter option is often preferable, since all the needed fluxes are readily available from EMD simulations, while the heat flux also necessitates computing partial enthalpies~\citep{de2013non} through post-processing~\citep{debenedetti1988_2}.

The energy flux is the first spatial moment of the time derivative of the energy density~\citep{baroni2020heat}:
\begin{align}\label{eq:energy flux md}
    \begin{split}
        \mathbf{J}_E(t) &= \int \mathbf{r} \dot{e}(\mathbf{r}, t) \dd[3]{r} \\
        &= \int \mathbf{r} \sum_{\ell=1}^N \left[\pdv{e(\mathbf{r},t)}{\mathbf{R}_\ell} \cdot \dot{\mathbf{R}}_\ell + \pdv{e(\mathbf{r},t)}{\mathbf{P}_\ell} \cdot \mathbf{f}_\ell \right] \dd[3]r,
    \end{split}
\end{align}
where $\mathbf{R}_\ell$ and $\mathbf{P}_\ell$ are the position and linear momentum of the $\ell$th atom, respectively, and $\mathbf{f}_\ell$ is force acting on it. There is no \emph{a priori} correct way to define the energy density of a condensed matter system. However, this ambiguity does not pose an issue, as this quantity is not directly measurable in experiments. Moreover, quantities dependent on it, such as thermal conductivity, remain unaffected by the precise expression of local densities. This concept is referred to as the \emph{gauge invariance of transport coefficients}~\citep{baroni2020heat, grasselli2021invariance}.

The explicit expression of the energy flux depends on the Hamiltonian of the system, which, for a classical system of $N$ particles, takes the general form:

\begin{align}\label{eq:hamiltonian force fields}
    \mathcal{H} = \sum_{\ell=1}^N \frac{P_\ell^2}{2M_\ell} + \mathcal{V}(\mathbf{R}_1, \ldots, \mathbf{R}_N),
\end{align}
where $\mathcal{V}$ denotes the potential energy. Given the gauge-invariance principle, any local breakdown of the energy density into atomic contributions is suitable for computing the energy flux~\citep{baroni2020heat, grasselli2021invariance}. A valid choice is thus
 
\begin{align}
    e(\mathbf{r}, t) = \sum_{\ell=1}^N \epsilon_\ell(t) \delta(\mathbf{r} - \mathbf{R}_\ell(t)),
\end{align}
where each atomic energy, $\epsilon_\ell$, is concentrated on the respective atom and can be expressed as:

\begin{align}\label{eq:atomic energy}
    \epsilon_\ell = \frac{P_\ell^2}{2M_\ell} + \mathcal{V}_\ell,
\end{align}
with the first term representing atomic kinetic energy and the second indicating the portion of potential energy assigned to the $\ell$th atom, such that $\mathcal{V}=\sum_\ell \mathcal{V}_\ell$. These atomic energies are phase-space variables dependent on time through atomic positions and momenta. Consequently, the energy flux in Eq.~\eqref{eq:energy flux md} becomes 

\begin{align}\label{eq:energy flux md 2}
    \mathbf{J}_E(t) = \sum_{\ell=1}^N \left[\epsilon_\ell \dot{\mathbf{R}}_\ell - \sum_{\ell'=1}^N \pdv{\mathcal{V}_{\ell'}}{\mathbf{R}_\ell} \cdot \dot{\mathbf{R}}_\ell \left( \mathbf{R}_\ell - \mathbf{R}_{\ell'} \right) \right].
\end{align}
This equation is well-defined in PBC, as it solely depends on interatomic distances calculated using the minimum-image convention. As such, the formula proves suitable for MD simulations involving bulk systems. 

The concepts discussed thus far are applicable to both empirical force fields (FFs) and MLPs. In the case of the latter, it is possible to generate MD simulations of \emph{ab initio} quality to sample the relevant fluxes, including the energy flux of Eq.~\eqref{eq:energy flux md 2}, thus leading to the computation of $\kappa$. This approach becomes especially significant for glasses, where, as previously mentioned, challenges arise in lattice-dynamical calculations involving MLPs. 
While for semi-empirical FFs assigning the atomic energies in Eq.~\eqref{eq:atomic energy} might seem reasonable and simple, especially in the case of pairwise potentials, it is less so in the case of MLPs. MLPs typically rely on a decomposition of the global target quantity into local, atom-centered contributions. This approach offers several advantages, such as computational scalability and the possibility of retrospective interpretation of patterns of local contributions, provided that local predictions are sufficiently ``rigid'', i.e.,~that their value is robust under perturbations of the training set and/or model parameters~\citep{chong2023robustness}. 

\section{Results}

In this Section we first develop, for demonstration purposes, a dataset for the paradigmatic amorphous solid vitreous silica (v-SiO\textsubscript{2}) based on a referece semi-empirical FF. We train different MLPs on this dataset, and analyze their ability in reproducing thermal conductivity calculations. We then discuss the current limitations of lattice dynamics in thermal conductivity calculations with MLPs, and how these are absent in EMD simulations. We finally apply GKMD to an amorphous alloy useful in the context of solid-state electrolytes.

\subsection{Toy model of vitreous silica}

\begin{figure*}[htb]
    \centering
    \includegraphics[width=2\columnwidth]{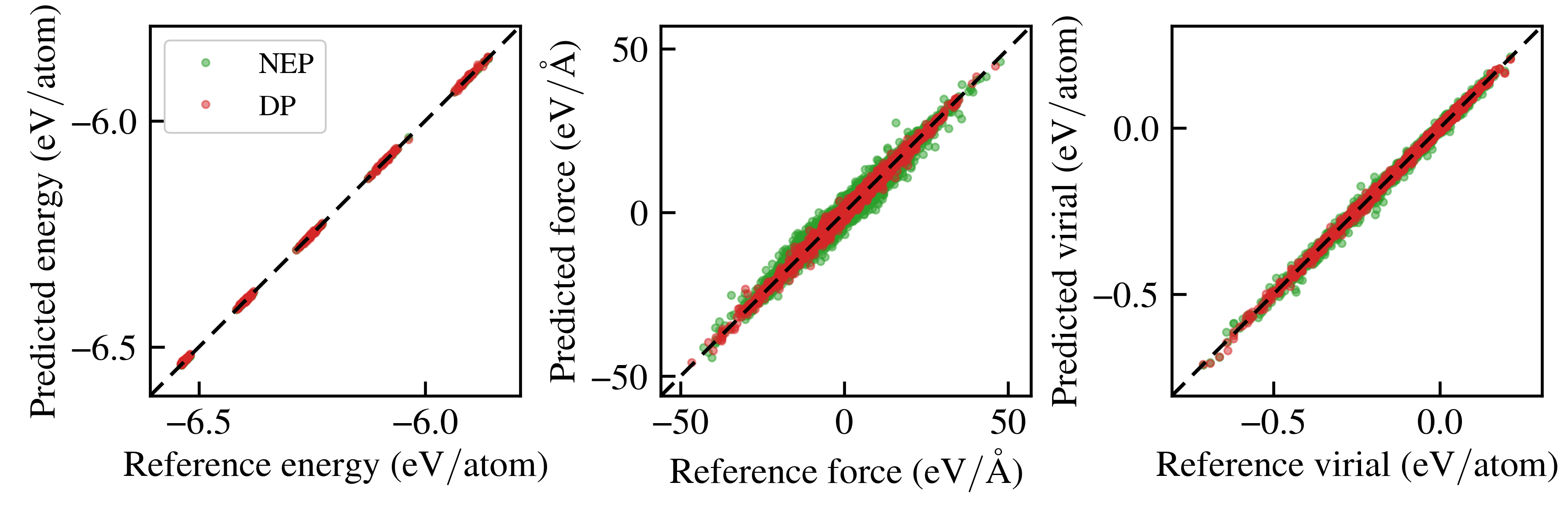}
    \caption{Parity plots of the committee of MLPs trained on the vitreous silica Tersoff dataset.}
    \label{fig:parity deepmd}
\end{figure*}

The reference potential is a Tersoff FF developed in~\cite{munethoh2007interatomic} as a short-range potential able to describe the amorphous phase of silica with relatively good accuracy. We choose a short-range reference potential in order to keep the model simple; moreover, the MLP to be trained is rigorously short-range, even if strategies to incorporate \textit{bona fide} long-range interactions exist and are being developed~\citep{grisafi2019incorporating,xie2020incorporating,ko2021fourth,veit2020predicting,zhang2022deep,huguenin2023physics}. 
The dataset is generated starting from 100 independently quenched glassy samples, equilibrated at 10 different temperatures ranging from 300\,K to 7000\,K. From each of the 100 simulations, 500 uncorrelated configurations are drawn once every 10\,ps. The dataset thus comprises 50,000 glassy and molten configurations. Further details can be found in the Supplementary Material.

We trained a committee of four deep neural network DeePMD potentials (DPs)~\citep{wang2018deepmd,zeng2023deepmd,lu2021pflops} smooth edition~\citep{zhang2018end} and a neuroevolution potential (NEP)~\citep{fan2021neuroevolution} to reproduce energies, forces and virials of the dataset. All the MLPs have a cutoff of {3\,\AA}, and are trained for more than 400 epochs to ensure the proper minimization of the loss function. The optimization of the loss function for the DPs is performed through the Adam stochastic gradient descent~\citep{kingma2017adam}, while for the NEP via a genetic algorithm~\citep{fan2021neuroevolution}. Each model within the DP committee differs from the others based on the initial random seed used in the minimization. The NEP features instead a different architecture and slightly different descriptors with respect to DPs~\citep{fan2021neuroevolution}.

We assess the performance of these models on a validation set, which consists of an additional 2500 configurations produced using the same methodology as the training set. Fig.~\ref{fig:parity deepmd} presents parity plots comparing the reference and predicted potential energies, forces, and virials for the four DP models in the committee. All the DPs exhibit comparable accuracy. The root mean square errors (RMSEs) and relative error with respect of the standard deviation (in parentheses), are $2.86 \pm 0.09\,\text{meV}/\text{atom}$ for the energy ($1\%$), $237.04 \pm 0.14\,\text{meV}/\text{\AA}$ for the force ($8\%$), and $6.96 \pm  0.04\,\text{meV}$ for the virial ($5\%$).
In Fig.~\ref{fig:parity deepmd} the performance of only one model of the DP committee is shown (red) for simplicity.
The NEP features RMSEs and relative errors of $0.9\,\text{meV}/\text{atom}$ for the energy ($< 1\%$), of $311\,\text{meV}/\text{\AA}$ for the force ($11\%$), and of $13\,\text{meV}$ for the virial ($8\%$).  

\subsection{Lattice dynamics vs molecular dynamics}

Subsequently, we test the models' performance in replicating the reference thermal conductivity. We compute the thermal conductivity of vitreous silica as determined by Eq.~\eqref{eq:gk kappa}. To do so, we sample the energy flux and the atomic mass fluxes through EMD simulations on a system with 648 atoms. Given that the Tersoff FF is a many-body potential, particular care needs to be taken when computing the energy flux~\citep{fan2015force}. A correct implementation of the Tersoff energy flux can be found in \textsc{gpumd}~\citep{fan2022gpumd}. As reference, we carried out a 1\,ns-long EMD simulation at 300\,K using \textsc{gpumd}, collecting the energy and mass fluxes every 5\,fs. Further computational details can be found in the Supplementary Material. We employ the same trajectory to sample these quantities using the committee of DPs and the NEP, utilizing the DeePMD and NEP implementations in \textsc{lammps}~\citep{LAMMPS,tisi2021heat,fan2022gpumd}. Using the same trajectory to test the models allows us to isolate the effects of gauge invariance, eliminating subtle differences that may arise when employing independent EMD runs to sample different trajectories. The thermal conductivity is finally obtained through cepstral analysis of the fluxes' time series~\citep{ercole2017accurate,bertossa2019theory}, as implemented in \textsc{SporTran}~\citep{ercole2022sportran}. The total energy decomposition into local contributions is in general different for different models, since the latter is a task arbitrarily performed by the ML algorithm. Thus, also the fluxes differ among the pool of models. Nevertheless, what must remain consistent is the physical observable, namely, the thermal conductivity, as commanded by gauge invariance~\citep{grasselli2021invariance}.

\begin{figure}[htb]
    \centering
    \includegraphics[width=\columnwidth]{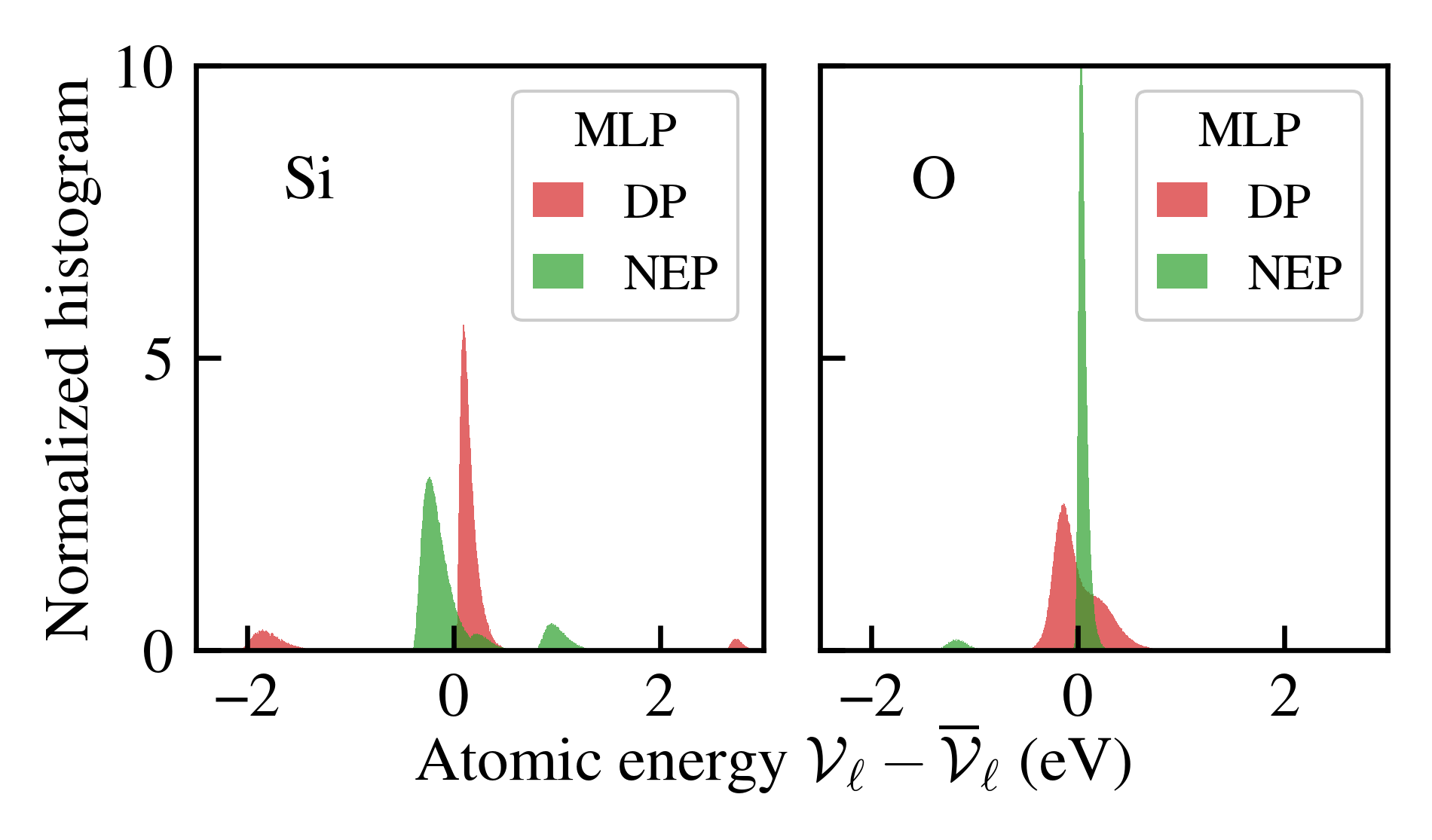}
    \caption{Distribution of the atomic energies of Si (left) and O (right) for the committee of MLPs.
    Simulations are carried out at $T=300\,\text{K}$ and ambient pressure.}
    \label{fig:committee 1}
\end{figure}

In Fig.~\ref{fig:committee 1}, we show the distributions (normalized histograms) of the deviation of the atomic energies per species with respect to their average value, sampled along the trajectory, as computed by the available MLPs. The DPs feature similar distributions of such quantity, so only one member of the committee is shown, while the NEP results are quite different. Nonetheless, even within the DP committee, the average values of the atomic energies are significantly different. In particular, the set of mean energies for the Tersoff FF, the DPs, and the NEP is reported in Tab.~\ref{tab:avg energies}. 
\begin{table}[htb]
    \centering
    \begin{tabular}{cccc}
        \toprule
        \; Model \; & \; $ \overline{\mathcal{V}}_\mathrm{Si}$ (eV) \; & \; $\overline{\mathcal{V}}_\mathrm{O}$ (eV) \; & \; $\overline{\mathcal{V}}_\mathrm{Si} + 2\overline{\mathcal{V}}_\mathrm{O}$ (eV) \; \\
        \midrule
        Tersoff & $-9.857$ & $-4.929$ & $-19.715$ \\
        DP1 &  $-5.413$ & $-7.155$ & $-19.723$ \\
        DP2 &  $-3.211$ & $-8.256$ & $-19.724$ \\
        DP3 &  $-4.606$ & $-7.555$ & $-19.715$ \\
        DP4 &  $-3.383$ & $-8.170$ & $-19.723$ \\
        NEP & $-7.833$ & $-5.941$ & $-19.716$ \\
        \bottomrule
    \end{tabular}
    \caption{Average atomic energies for the different models. The last column is the stoichiometry-weighted sum of average energies. It coincides for all the models, as it should, it being a physical observable. The discrepancy among this quantity's reported values is well below the threshold of chemical accuracy ($\sim 40\,$meV).}
    \label{tab:avg energies}
\end{table}

\begin{figure}[htb]
    \centering
    \includegraphics[width=\columnwidth]{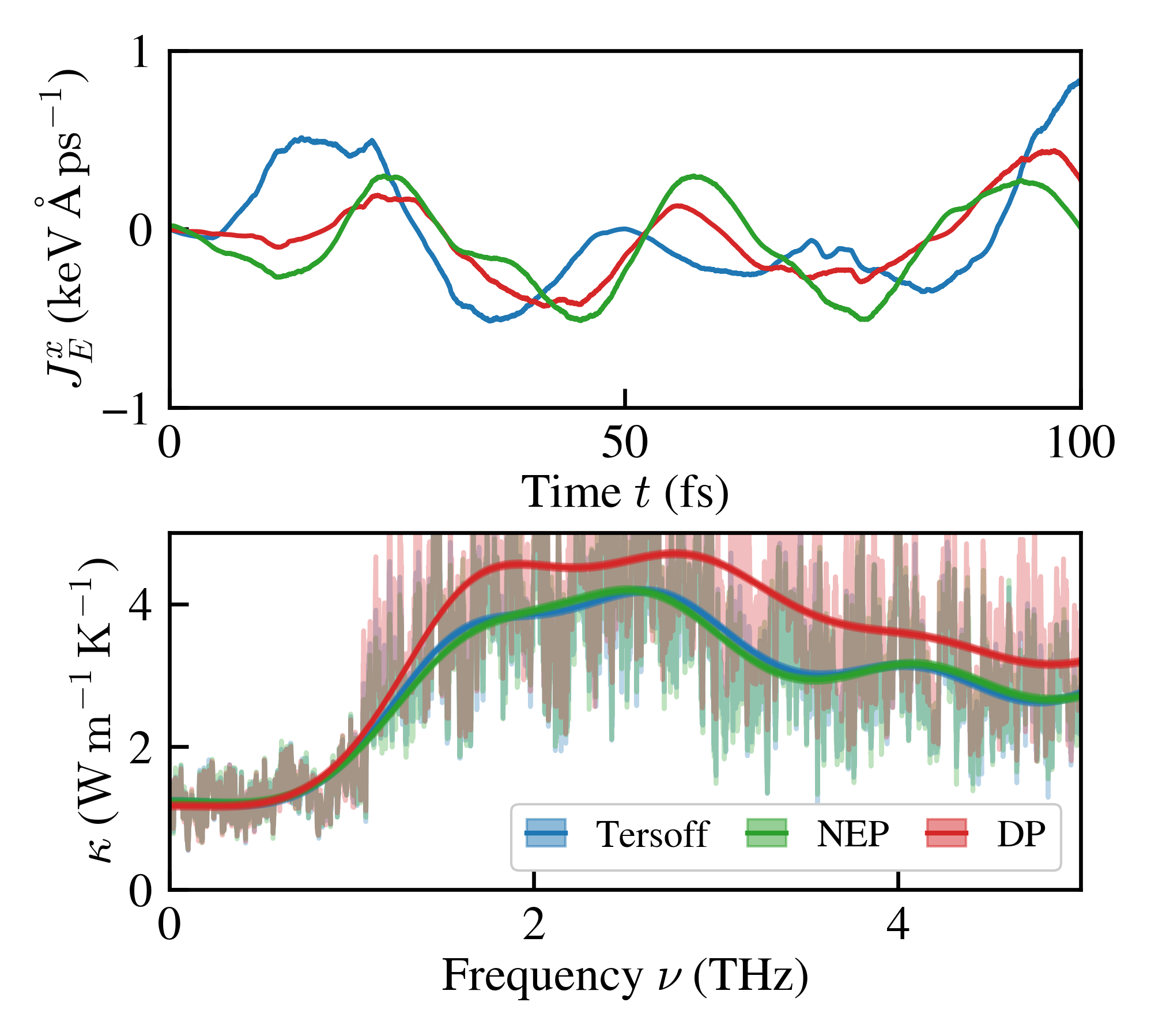}
    \caption{Thermal conductivity of the MLPs committee. Simulations are carried out at $300\,\text{K}$ and ambient pressure. Top panel: the reference (Tersoff) energy flux and the DP committee ones are different, in general. Bottom panel: cepstral analysis of the fluxes' time series. The thermal conductivity, given by the zero-frequency value of the curve, is the same for different models, as commanded by gauge invariance.}
    \label{fig:committee 2}
\end{figure}

We stress again that such a difference is expected, as the atomic energies are not target quantities in the learning scheme, and only the total energy is physically observable.
The difference in atomic energies predicted by the different models is translated into a difference in energy fluxes, as shown in the top panel of Fig.~\ref{fig:committee 2}. Crucially, the thermal conductivity, as represented by the zero-frequency value of the fluxes' power spectral density~\citep{ercole2017accurate,bertossa2019theory}, remains unchanged, as reported in the bottom panel of Fig.~\ref{fig:committee 2}.

\begin{figure}[htb]
    \centering
    \includegraphics[width=\columnwidth]{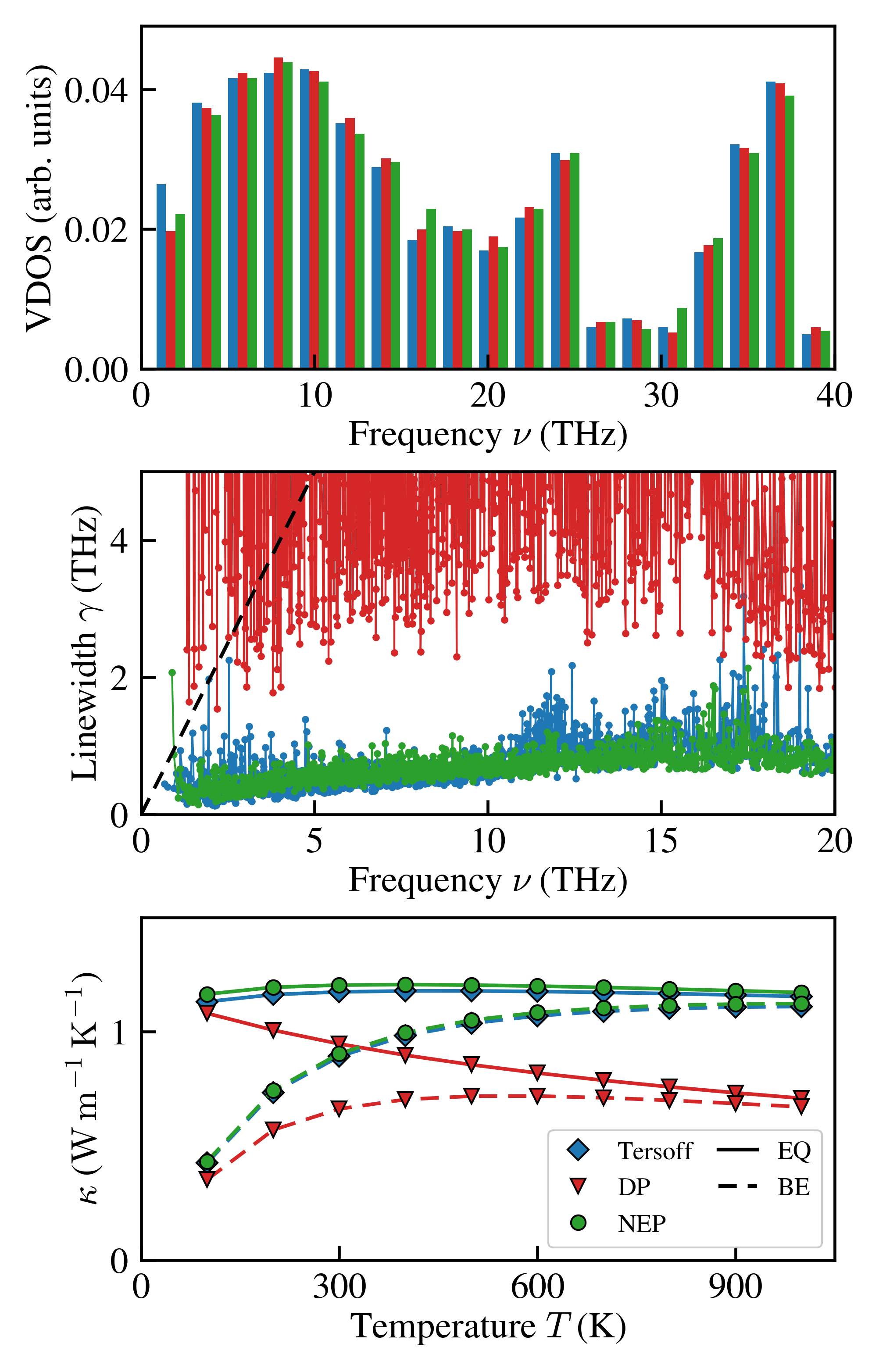}
    \caption{Results based on lattice dynamics of vitreous silica computed with the Tersoff FF, the DP, and the NEP. Upper panel: vibrational density of states computed with the three models. Central panel: normal-mode linewidths. Lower panel: QHGK thermal conductivity. EQ stands for the classical equipartition, while BE for the Bose-Einstein occupation.}
    \label{fig:lattice dynamics}
\end{figure}

We now assess the models' performance in reproducing quantities relevant for lattice-dynamical calculations. We compute the second and third-order interatomic force constants with \textsc{lammps} and obtain normal-mode frequencies and linewidths using {$\kappa$ALD$o$}~\citep{barbalinardo2020efficient}. We employ these quantities to compute the QHGK thermal conductivity with the Tersoff FF, a DP, and the NEP. Fig.~\ref{fig:lattice dynamics} showcases the results, including the vibrational density of states (VDOS), normal-mode linewidths, $\gamma$, and QHGK thermal conductivity. While the VDOS outcomes in the upper panel exhibit agreement, the same cannot be said for anharmonic normal-mode linewidths (central panel), which appear to be accurately captured only by the NEP. In contrast, the DP results considerably overestimate the linewidths, even surpassing the quasi-harmonic regime, denoted by the dashed black line. These disparities have direct implications for the computed QHGK thermal conductivities, as evident in the lower panel of Fig.~\ref{fig:lattice dynamics}. Specifically, while the NEP and Tersoff results align with each other, the DP results do not share this agreement. 

\begin{figure}[htb]
    \centering
    \includegraphics[width=\columnwidth]{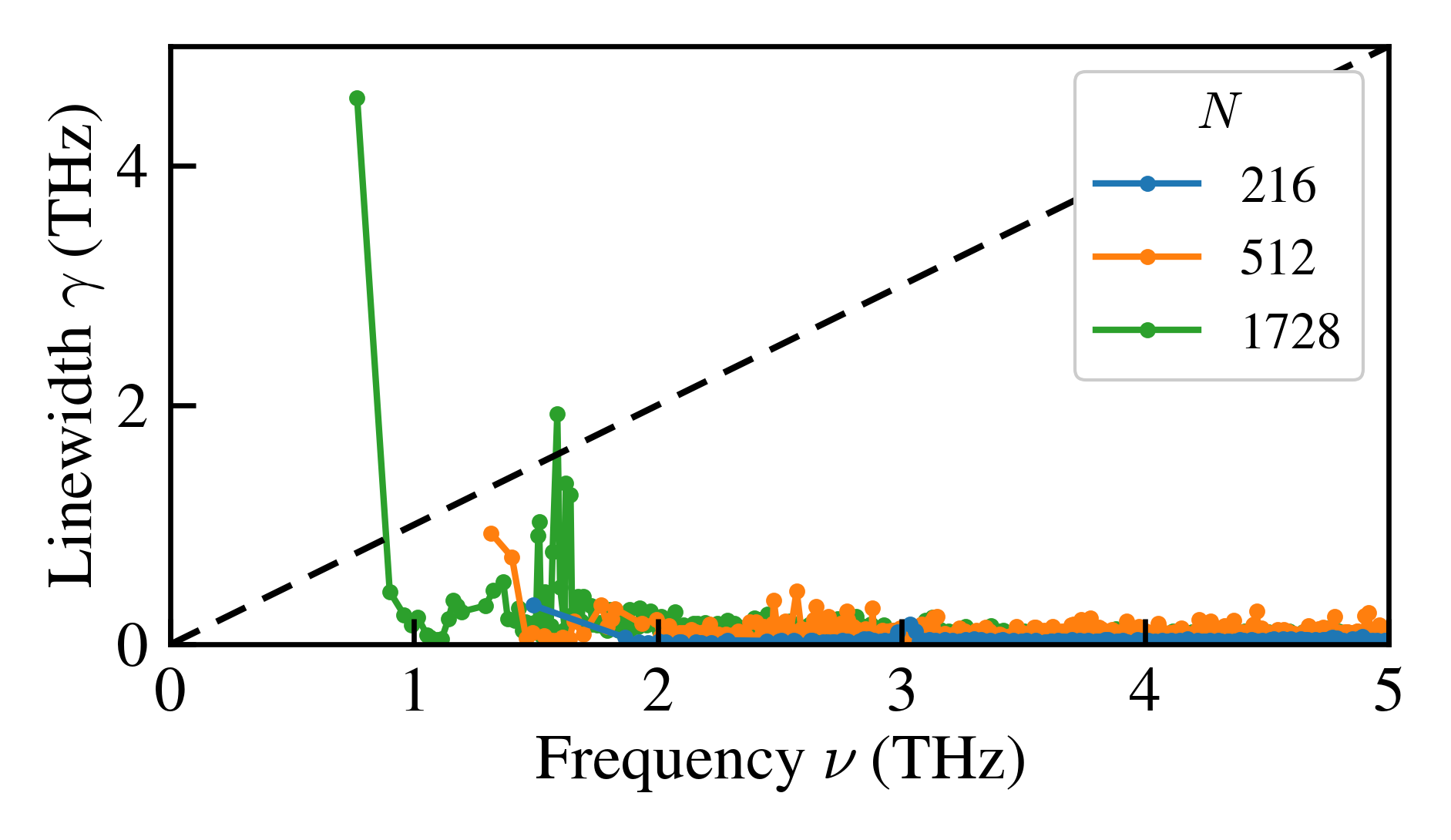}
    \caption{Anharmonic linewidths of a NEP model for v-Si at $300\,$K, for samples of different sizes.}
    \label{fig:asi linewidth}
\end{figure}

It is essential to recognize that apparently satisfactory performance of the NEP leads to potentially misleading outcomes. Upon closer examination of the central panel of Fig.~\ref{fig:lattice dynamics}, a notable issue emerges: at low frequencies (in this instance, the first available finite frequency) there are disproportionately large linewidths. This stark contrast with the expected decrease in linewidths as frequencies decrease, as dictated by the hydrodynamics of solids~\citep{griffin1968brillouin, fiorentino2023hydrodynamic}, precludes the hydrodynamic extrapolation the NEP results to obtain a size-converged value for bulk thermal conductivity~\citep{fiorentino2023hydrodynamic, fiorentino2023unearthing}. As a further proof that MLPs still present some issues with lattice dynamics, we compute the anharmonic linewidths of a more complex NEP for vitreous silicon (v-Si) developed in~\cite{wang2023quantum} and based on the dataset of~\cite{bartok2018machine}. As shown in Fig.~\ref{fig:asi linewidth}, increasing the size of the sample exacerbates the issue of the low-frequency peak in the linewidths, which hinders the possibility of carrying out size-converged thermal conductivity calculations~\citep{fiorentino2023hydrodynamic,fiorentino2023unearthing}. This issue persists even when manually imposing acoustic sum rules due to global translational invariance and symmetry under Cartesian-direction permutation.  

Due to these limitations, lattice-dynamical calculations prove to be of limited utility when employing MLPs. Consequently, the alternative approach is to resort to GKMD simulations, albeit at the cost of neglecting the quantum BE occupation. It is worth noting that in specific instances, like vitreous silica, this omission remains notable even at room temperature~\citep{lv2016nonnegligible,zhu2022effect}. Conversely, for other simple glasses like vitreous silicon, this concern does not apply~\citep{fiorentino2023hydrodynamic,wang2023quantum}. 

\subsection{Thermal conductivity of Li-intercalated silicon}

To exemplify the GKMD methodology within the context of MLPs, we conduct a thermal conductivity assessment of a PBE-accurate model describing amorphous LiSi, as developed by~\cite{fu2023unraveling}. This MLP was employed to investigate the properties of silicon, encompassing both crystalline and amorphous phases, as an anode material for Li-based electrolytes~\citep{fu2023unraveling}. Thermal transport plays a pivotal role in the design of electrolytes, especially for solid-state systems~\citep{feng2018thermal}. Indeed, an exceedingly low thermal conductivity can lead to excessive heat generation, especially during rapid charging processes, thereby posing the risk of critical incidents such as material melting or explosions. Moreover, the management of thermal dissipation is crucial to optimizing energy conservation and utilization, requiring a delicate equilibrium between minimizing heat dissipation and maximizing electric flux throughout the charging cycle~\citep{feng2018thermal}. In addition, in situations involving materials where ionic diffusion is not only expected but also a desired attribute, such as solid-state electrolytes, one is forced to use methods that do not rely on the presence of well-defined atomic equilibrium positions~\citep{pegolo2022temperature}.

We conduct EMD simulations on Li\textsubscript{$x$}Si\textsubscript{$1-x$} with varying concentration of intercalated lithium, denoted as $x$, under room temperature conditions ($T=300\,$K, $p=0\,$bar). Amorphous samples are prepared through a melt-quench-anneal procedure as outlined in~\cite{fu2023unraveling}. The samples so obtained are used as initial configurations for EMD simulations that sample the relevant fluxes in canonical runs~\citep{bussi2007canonical} carried out for 3\,ns at 300\,K.
The simulations are performed with \textsc{gpumd} using a NEP trained on the dataset of~\cite{fu2023unraveling}. Further details can be found in the Supplementary Material.

\begin{figure}[b]
    \centering
    \includegraphics[width=\columnwidth]{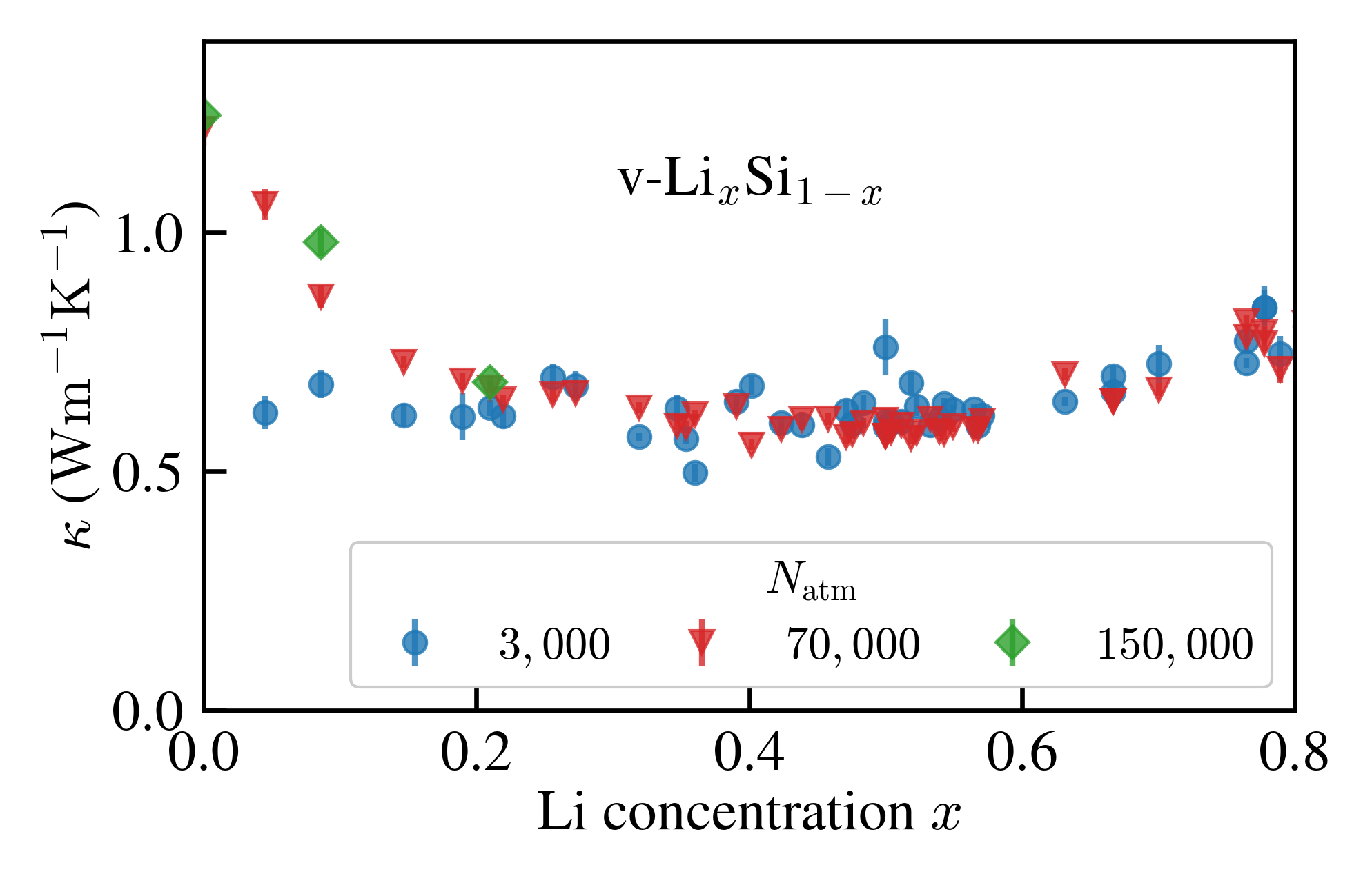}
    \caption{Thermal conductivity of v-Li\textsubscript{$x$}Si\textsubscript{$1-x$} as a function Li concentration at room-temperature conditions for different system sizes.}
    \label{fig:lisi kappa}
\end{figure}

Results for the thermal conductivity as a function of Li concentration are presented in Fig.~\ref{fig:lisi kappa}. Size effects are important for low Li concentrations, where the system is close to pure vitreous silicon. v-Si is known for being severely affected by size effects~\cite{fiorentino2023hydrodynamic} due to its high local order~\cite{bartok2018machine}. Thus, samples with 3,000 atoms are not converged in size, while 70,000 atoms appear to be enough, when compared to calculations on systems with 150,000 atoms.  

At higher Li concentrations, even 3,000 atoms are sufficient to achieve convergence. The behavior of $\kappa(x)$ features a minimum $x \approx 0.5$, followed by an increase for larger values of $x$. Pure v-Si features $\kappa=1.24\pm0.03\,\mathrm{W\,m^{-1}\,K^{-1}}$. This value can be compared with calculations~\citep{wang2023quantum} done on structure of similar size that employ a NEP trained on a dataset relying on PW91 DFT calculations~\citep{bartok2018machine}, rather than PBE. The results are compatible at high quenching rate, while we report a value which is $\approx 30$\% lower than the one of~\cite{wang2023quantum} at the lowest quenching rate. Experimental data on the thermal conductivity of v-Si severely depend on the experimental sample size, due to the high relevance of propagons in this material. At 300\,K, they can range from less than $1\,\mathrm{W\,m^{-1}\,K^{-1}}$~\cite{zink2006thermal} to around $4\,\mathrm{W\,m^{-1}\,K^{-1}}$~\cite{liu2009high, yang2010anomalously}. We are not aware of existing experimental data on the thermal conductivity of Li-intercalated amorphous silicon.

The qualitative behavior of the thermal conductivity as a function of Li concentration aligns with calculations on crystalline alloys~\citep{garg2011role}, where the distinctive U-shape of $\kappa(x)$ is interpreted in terms of phonon scattering due to isotopic mass disorder~\citep{tamura1983isotope} at the perturbative level. Conversely, in glasses the thermal conductivity reduction with increasing concentration of different-species atoms is mostly due to the increased localization of low- and mid-frequency vibrational modes, together with the broadening of the respective linewidths, which hinders their ability to transport heat~\citep{mode2021lundgren}.

\section{Conclusions}

In this work, we have reviewed the theory of thermal transport in amorphous solids, focusing on the role of MLPs as a tool to expedite and, in some cases, allow for thermal-transport characterization of glasses from atomistic simulations. We have build an example of MLP for vitreous silica able to reproduce the potential energy surface of a Tersoff FF, and used it to highlight some challenges still present when dealing with lattice dynamics, leaving Green-Kubo molecular dynamics as a valuable alternative to carry out calculations with MLPs. We then applied such methodologies to the Li-concentration dependence of the thermal conductivity of lithium-intercalated amorphous silicon, finding that $\kappa$ features a minimum at half concentration, coherent with analogous calculations made on other amorphous alloys and interpreted in terms of the localization of propagating modes due to chemical disorder. 

\section*{Data Availability Statement}

The data and scripts that support the plots and relevant results within this paper are available on Zenodo~\citep{pegolo2023data}.

\section*{Funding}
This work was partially supported by the European Commission through the \textsc{MaX} Centre of Excellence for supercomputing applications (grant number 101093374) and by the Italian MUR, through the PRIN project \emph{FERMAT} (grant number 2017KFY7XF) and the Italian National Centre from HPC, Big Data, and Quantum Computing (grant number CN00000013). FG acknowledges funding from the European Union's Horizon 2020 research and innovation programme under the Marie Sk\l{}odowska-Curie Action IF-EF-ST, grant agreement number 101018557 (TRANQUIL).

\newpage

\section*{Acknowledgments}

We thank A. Fiorentino and D. Tisi for fruitful discussions. We are grateful to S. Baroni for his valuable guidance.

\end{document}